\documentclass{article}
\usepackage{epsfig}
\begin{document}
\begin{center}
{{\bf \Large Research News --- Observation of 
oscillation phenomena in heavy meson systems
}}

\bigskip

B. Ananthanarayan$^a$, 
Keshav Choudhary$^b$,
Lishibanya Mohapatra$^b$\\
Indrajeet Patil$^c$,
Avinash Rustagi$^b$,
K. Shivaraj$^a$

\bigskip
{\small
$^a$ Centre for High Energy Physics, Indian Institute of Science,
Bangalore 560 012\\
$^b$ St. Stephen's College, Delhi 110 007\\
$^c$ Fergusson College, Pune 411 004\\ 
}
\medskip
\end{center}

\noindent{\bf Keywords:} Heavy mesons, oscillation phenomena 

\medskip

\begin{abstract}
We review the recent discoveries of 
rare oscillation phenomena in certain heavy neutral meson systems. 
\end{abstract}

We report on the recent observation of certain `matter-antimatter' oscillations
in neutral meson systems, which have long been theoretically predicted, but 
whose experimental verification has remained a great challange. These are the 
$D^0- \overline{D^0}$ oscillations discovered by the BABAR and BELLE 
Collaborations at the PEP-II and KEK-B facilities in the USA and Japan
respectively, and the $B_s^0 - \overline{B^0}_s$ oscillations discovered by the 
CDF Collaboration at the Fermi National Accelerator
Laboratory in the USA a little earlier.

Strongly interacting particles come in two varieties, baryons and mesons.
For instance, protons and neutrons are baryons and are made up of
three constituent quarks.  Among mesons, a familiar example is the
pion which is the transmitter of the internucleon force.  It is 
made up of a quark and anti-quark pair.  These particles are
the residue of the strong interactions that bind quarks through the
exchange of gluons and the microscopic theory that describes these
interactions is known as quantum chromodynamics.  

It has now been experimentally established that
quarks come in six flavours, of which all but the u- quark
are unstable due to the presence of the weak interactions.  The latter
is responsible for the decay of a free neutron, and is the only force
that allows for the change in particle type.  In Table 1, we
present the list of the quarks and their electric charges.
\vspace{0.3cm}
\begin{center}
\begin{tabular}{|c|c|c|}
\hline
$Quark$&$Symbol$&$Charge$\\
\hline
Up&u&+2/3\\
Down&d&-1/3\\
Charm&c&+2/3\\
Strange&s&-1/3\\
Top&t&+2/3\\
Bottom&b&-1/3\\
\hline
\end{tabular}  
\end{center}
\vspace{0.3cm}
It may be seen that we can form only four kinds of neutral mesons and
their respective anti-particles out of
this set, namely the $K$-system ($\overline{s} d, s \overline{d}$), 
the $B$-system ($\overline{b} d, b \overline{d}$), the 
$B_s$-system ($\overline{b} s, b \overline{s}$), 
and the $D$-system ($\overline{c} u, c \overline{u}$). The top-quark, 
on the other hand is so massive and consequently so unstable that 
it does not have time to form mesons before its decay.  
There are specific phenomena associated with
these systems, which provide
a sensitive test of the `standard model', the theory of electroweak and strong
interactions and of quantum mechanical phenomena in general.  
Of these phenomena, we concentrate on a specific one known as
oscillation between a meson and the corresponding anti-meson.
The reason for the oscillation is the fact that the weak interactions do
not conserve particle type.   Recalling that the weak interaction
at leading order in the interaction strength violates `flavour'
conservation, {\it viz.} we can have transitions of the kind
$d\to u,\, s\to u, c\to s, c\to d, b\to c, b\to u$, which leads
to the decay of these particles. 
It is important to note that not all allowed transitions are
equally likely.  For instance,
the transition $c \to s \ (\propto \cos\theta_c)$ 
is favoured over $c \to u \ (\propto \sin\theta_c)$, since $\theta_c$
which is known as the Cabibbo angle is 
rather small ($\approx\, 13.1^0$). A transition that 
is proportional to $\cos\theta_c$ is referred as `Cabibbo favoured' while 
a transition that is proportional to $\sin\theta_c$ is referred as `Cabibbo
suppressed'.

The same interaction at higher order
can lead to the formation of virtual intermediate states which can
then transform by violating flavour conservation by one more unit,
thereby leading effectively to $\Delta F=2, \, F=S,\, C,\, B$(here S, C, B are
corresponding quantum numbers) interactions. Such interactions, can 
turn, e.g., a $K^0$ meson into a $\overline{K^0}$ before the 
particle decays.   This is the phenomenon of oscillation also known as mixing.
This was detected in the $K$- system about 50 years ago 
~\cite{Phys.Rev.103.1901, Phys.Rev.103.1904}, and in the $B$-system 
about 20 years 
ago~\cite{Phys.Lett.B186.247, Phys.Lett.B192.245}. Here we
review the brand new developments in this field with the 
reported observations of oscillation phenomena in the $D$-meson system 
earlier this year, and the prior discovery in $B_s$-system in 2006.
The reason for the long delay in the observation of the former is
intimately related to the structure of the standard model; 
the small masses of the b- and s- quarks render the
mixing in the $D$- system to be very small. The
corresponding delay in the detection of oscillation
phenomena in the $B_s$ system is linked to the very high
oscillation frequency.

The discovery of oscillation phenomena in the $D$- meson system
commenced with the announcement by the BABAR Collaboration
~\cite{Phys.Rev.Lett.98.211802}
at the PEP-II storage ring at Stanford Linear Accelerator Center(SLAC).
This was followed by the announcement of the
BELLE Collaboration at the KEK-B in Japan~\cite{Phys.Rev.Lett.98.211803}. 
These 
discoveries have taken place in what is known as B- factories.
These experiments have detected specific final states of the decays
of the mesons which are sensitive to mixing; they have now ruled out
the absence of mixing at `$3.9 \, \sigma$' (BABAR) and 
`$3.2\, \sigma$' (BELLE) level, respectively.  
The CDF Collaboration in April, 2006 announced the discovery of
oscillation phenomena in the $B_s$ system~\cite{Phys.Rev.Lett.97.242003}, 
which confirmed the prior observations of the D0 
Collabaration~\cite{Phys.Rev.Lett.97.021802} also at the Fermilab Tevatron,
which for technical reasons could  not be considered a `discovery'.

The results are summarized in terms of specific parameters characterizing
the mixing phenomena.  Consider starting out with the
two states $D^0$ and $\overline{D^0}$ in the absence of mixing.  In
the presence of mixing, they would propagate effectively as two states,
which are linear combinations of $D^0$ and $\overline{D^0}$ as
dictated by the principles of quantum mechanics,
with masses $M_1$, $M_2$ and inverse lifetimes or `widths' denoted 
by $\Gamma_1$, $\Gamma_2$, keeping in mind that  
that the weak interaction is the 
force that leads to mixing as well as to decay.
In the literature~\cite{hep-ph/0703245,hep-ph/0703235}, 
the mixing phenomena are parametrized  
in terms of two dimensionless parameters defined as,
\begin{eqnarray*}
& \displaystyle x={2(M_1-M_2)\over \Gamma_1+\Gamma_2}, \ \ y={\Gamma_1-
\Gamma_2\over \Gamma_1+\Gamma_2}.
\end{eqnarray*}
In the absence of mixing $x$ and $y$ would vanish identically. 
It may be recalled that it is only in the neutral meson systems that
thus far the phenomenon of CP violation has been observed, where C stands for
charge conjugation or particle-antiparticle symmetry, and P stands for
parity or invariance under reflection,  with CP denoting the
operation C following the operation P. It turns out
that many of the CP violating 
observations are intimately linked to the phenomena
of mixing which have been observed 
in the $K$- and $B-$ systems.  However, in the two most
recent discoveries, this phenomenon is essentially absent, thereby
simplifying the analysis of the oscillation phenomena.

The BABAR detector at PEP-II storage rings at SLAC announced the evidence for 
$D^0 - \overline{D^0}$ mixing in $D^0 \to K^+ \pi^-$ decays from 
$348 \ {\rm fb^{-1}}$ of $e^- e^+$ colliding beam data recorded near 
$\sqrt{s} = 10.6\  GeV$ ( $\sqrt{s}$ is the total energy in the Center of 
Mass frame of $e^-$ and $e^+$, and ${\rm fb}^{-1}$ stands for
inverse femtobarn, a standard unit of luminosity).

The BABAR Collaboration bases its report on the observation of 
`right-sign' (RS) and `wrong-sign' (WS) events in the final state $K \pi$. The
corresponding decays are shown in Figs. 1 and 2.
\begin{figure}[ht]
\centering
\includegraphics[width=12cm]{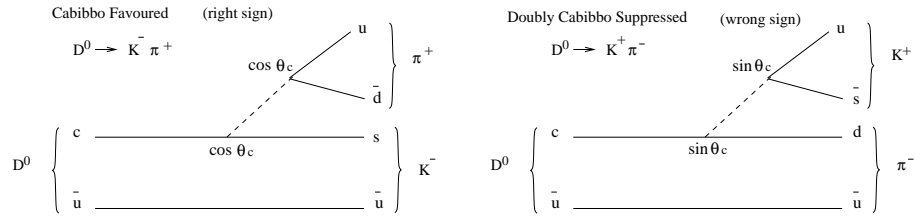}
\caption{The figure shows the RS and WS events, without the 
oscillation $D^0 \to \overline{D^0}$}
\end{figure}   

\begin{figure} [ht]
\centering
\includegraphics[width=12cm]{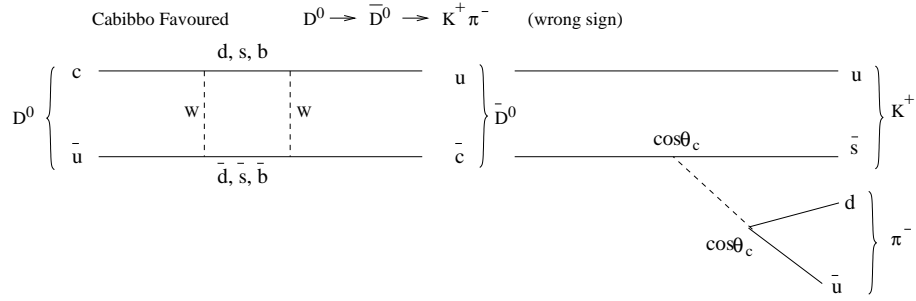}
\caption{The figure shows the Cabibbo favoured decay of $D^0$ meson, after
oscillation, going to $K^+ \pi^-$.}
\end{figure}

The BABAR Collaboration focussed on the WS decay $D^0\to K^+ \pi^-$,
which bears the imprint of mixing.
The desired `wrong state' $K^+ \pi^-$ can be obtained in two ways: a) through 
Doubly Cabbibo Suppressed (DCS) decay and b) through Cabibbo Favoured (CF) 
decay after oscillating to $\overline{D^0}$. 
Each of these two contributions to the WS decays provides a different time
dependence to the decay rates which the experiment measures.
$D^0$ and 
$\overline{D^0}$ are distinguished by their production in the decay 
$D^{*+} \to \pi^+_s D^0$ where $\pi^+_s$ is referred as `slow pion'. Here 
the $D^{*+}$ is an excited state of $D^+$. 
Thus, the RS and WS decays are:
\begin{eqnarray*}
D^{*+} \to \pi^+_s D^0 &\to& \pi^+_s K^- \pi^+ \hspace{1cm} {\rm (RS)} \\
                       &\to& \pi^+_s K^+ \pi^-  \hspace{1cm}{\rm (WS)}
\end{eqnarray*}  

In RS decay, the $\pi^+_s$ and kaon have opposite charges, while in WS decay, 
the $\pi^+_s$ and kaon have same charge. Time dependence of the WS decay 
rate is used to separate DCS decays from $D^0 - \overline{D^0}$ mixing. BABAR 
Collaboration analysed $1141500 \pm 1200$ RS signal events and 
$4030 \pm 90$ WS signal events. The data was analysed in terms of parameters 
$x'$ and $y'$, where they are related to $x$ and $y$ by a strong 
phase ($\delta$) associated with the rescattering of the $\pi K$ pair
in the final state and the relations are, 
\[x' = x \ \cos\delta + y \ \sin\delta \] 
\[y' = y \ \cos\delta - x \ \sin\delta \]
The measured values are,
\begin{eqnarray*}
y'&=& [9.7 \pm 4.4({\rm stat}) \pm 3.1({\rm syst})] \times 10^{-3}  \\
{x'}^2 &=& [-0.22 \pm 0.30({\rm stat}) \pm 0.21({\rm syst})] \times 10^{-3} \\
\end{eqnarray*}
This result is inconsistent with the no-mixing hypothesis with a 
significance of 3.9 $\sigma$.

The BELLE Collaboration based their results 
on $540 \ {\rm fb^{-1}}$ of data recorded at KEK-B asymmetric-energy collider. 
Their idea is based on the fact that if there is oscillation then there will 
be a time difference between
decays $D^0 \to K^- \pi^+ $, $D^0 \to K^+ K^-$, and $D^0 \to \pi^+ \pi^-$. 
Using the fact that there was no
evidence for CP violation, 
BELLE was able to measure the lifetime difference 
of $D^0$ mesons between these 
decays to be 

\begin{eqnarray*}
& \displaystyle
y_{CP}=(1.31\pm 0.32 ({\rm stat}) \pm 0.25({\rm syst}))\ \times 10^{-2}, & \\
& \displaystyle
y_{CP}\equiv {\tau(K^-\pi^+)\over \tau(K^+K^-)}-1,
\end{eqnarray*} 
where $\tau(K^-\pi^+)$ and $\tau^(K^+K^-)$ are the lifetimes 
$D^0\to K^-\pi^+$ and $D^0\to K^+K^-$ respectively,  
and $y_{CP}$ equals $y$ in the limit of CP conservation
Thus the BELLE result differs from 
zero by 3.2 $\sigma$. Non-zero lifetime difference between different decay 
modes represents clear evidence for the transformation of $D^0$ particle 
into its anti-particle.
 
Therefore, both BABAR and BELLE are able to show that there is oscillation in 
$D$- systems and they did not observe any CP violation. It is partly because of 
the limitation on the sensitivity of the experiments. The new experiments with 
higher accuracy could reveal the fate of CP violation in $D$- systems.

Turning now to the oscillation phenomena in the $B_s$- system, 
these have been observed by the CDF Collaboration by considering the 
$B_s$ hadronic decays
$\overline{B^0}_s \to D^+_s \pi^-$ and   $D^+_s \pi^- \pi^+ \pi^-$, 
and the semileptonic decays $\overline{B^0}_s \to D_s^{+(*)} l^- 
\overline{\nu}_l , l = e \ {\rm or}
 \ \mu $. Using the probability density functions 
that describe the measured time development of $B_s$ mesons that decay with 
same or opposite flavour as their flavour at production, they have measured
the value of $\Delta m_s$ (oscillation frequency). The 
measured $\Delta m_s = 17.77 \pm 0.10({\rm stat})\pm 
0.07({\rm syst}) \ {\rm  ps^{-1}}$ with more than $5 \sigma$ confidence level. 

In summary, we have reviewed the recent experimental discoveries of oscillation
phenomena in the two neutral systems where the effects are hardest to observe. 
In the coming years the oscillation parameters are likely to be well-measured
and CP violation may be seen. A large signal for CP violation in
these neutral meson systems would 
imply the presence of interactions beyond those described by the standard model.

\end{document}